\def\pmb#1{\setbox0=\hbox{#1}%
   \kern-.025em\copy0\kern-\wd0
   \kern.05em\copy0\kern-\wd0
   \kern-0.025em\raise.0433em\box0}
\def\gta{\mathrel{{\lower 3pt\hbox{$\mathchar"218$}}\hskip-8pt
   \raise 2pt\hbox{$\mathchar"13E$}}}
\def\lta{\mathrel{{\lower 3pt\hbox{$\mathchar"218$}}\hskip-8pt
   \raise 2pt\hbox{$\mathchar"13C$}}}
\def\half{{\scriptstyle{1\over2}}}
\def\dagg{\phantom{\dagger}}            

\documentclass{kapproc} 

\usepackage{procps}

\usepackage[dvips]{graphicx}

\upperandlowercase

\kluwerbib 

\normallatexbib 

\font\tenrm=cmr10 
\begin{document}

\articletitle{Stripe-like Inhomogeneities, Coherence, and the Physics of
the High $T_c$ Cuprates} 


\author{J. Ashkenazi}
\affil{Physics Department, University of Miami, P.O. Box 248046, Coral
Gables, FL 33124, U.S.A.\\} 
\email{jashkenazi@miami.edu}

\begin{abstract}
The carriers in the high-$T_c$ cuprates are found to be polaron-like
``stripons'' carrying charge and located in stripe-like inhomogeneities,
``quasi-electrons'' carrying charge and spin, and ``svivons'' carrying
spin and some lattice distortion. The anomalous spectroscopic and
transport properties of the cuprates are understood. The stripe-like
inhomogeneities result from the Bose condensation of the svivon field,
and the speed of their dynamics is determined by the width of the
double-svivon neutron-resonance peak. The connection of this peak to the
peak-dip-hump gap structure observed below $T_c$ emerges naturally.
Pairing results from transitions between pair states of stripons and
quasi-electrons through the exchange of svivons. The pairing symmetry is
of the $d_{x^2-y^2}$ type; however, sign reversal through the charged
stripes results in features not characteristic of this symmetry. The
phase diagram is determined by pairing and coherence lines within the
regime of a Mott transition. Coherence without pairing results in a
Fermi-liquid state, and incoherent pairing results in the pseudogap
state where localized electron and electron pair states exist within the
Hubbard gap. A metal-insulator-transition quantum critical point occurs
between these two states at $T=0$ when the superconducting state is
suppressed. An intrinsic heterogeneity is expected of superconducting and
pseudogap nanoscale regions. 
\end{abstract}


\section{Introduction}

It is suggested that the anomalous physics of cuprates, including the
existence of high-$T_c$ superconductivity, can be understood as the
result of a behavior typical of their structure within the regime of a
Mott transition. Theoretical calculations, and a variety of experimental
data, support the assumption that their microscopic structure, within
this regime, is often characterized by dynamic stripe-like
inhomogeneities. 

To study this regime, a combination of large-$U$ and small-$U$ orbitals
is considered, where major aspects within the CuO$_2$ planes are
approached by the $t$--$t^{\prime}$--$J$ model. The large-$U$ electrons,
residing in these planes, are treated by the ``slave-fermion'' method
\cite{Barnes}. Such an electron in site $i$ and spin $\sigma$ is created
by $d_{i\sigma}^{\dagger} = e_i^{\dagger} s_{i,-\sigma}^{\dagg}$, if it
is in the ``upper-Hubbard-band'', and by $d_{i\sigma}^{\prime\dagger} =
\sigma s_{i\sigma}^{\dagger} h_i^{\dagg}$, if it is in a Zhang-Rice-type
``lower-Hubbard-band''. Here $e_i^{\dagg}$ and $h_i^{\dagg}$ are
(``excession'' and ``holon'') fermion operators, and
$s_{i\sigma}^{\dagg}$ are (``spinon'') boson operators. These auxiliary
operators have to satisfy the constraint: $e_i^{\dagger} e_i^{\dagg} +
h_i^{\dagger} h_i^{\dagg} + \sum_{\sigma} s_{i\sigma}^{\dagger}
s_{i\sigma}^{\dagg} = 1$. 

An auxiliary space is introduced within which a chemical-potential-like
Lagrange multiplier is used to impose the constraint on the average.
Physical observables are projected into the physical space by expressing
them as combinations of Green's functions of the auxiliary space. Since
the time evolution of Green's functions is determined by the Hamiltonian
which obeys the constraint rigorously, it is not expected to be violated
as long as justifiable approximations are used. 

The reader is referred to previous publications by the author
\cite{Ashk1,Ashk2} which include technical details concerning the work
discussed here. 

\section{Auxiliary Fields}

Uncoupled auxiliary fields are considered at the zeroth order, where the
spinon field is diagonalized by applying the Bogoliubov transformation
for bosons \cite{Ashk3}: 
\begin{equation}
s_{\sigma}^{\dagg}({\bf k}) =
\cosh{(\xi_{\sigma{\bf k}})} \zeta_{\sigma}^{\dagg}({\bf k}) +
\sinh{(\xi_{\sigma{\bf k}})} \zeta_{-\sigma}^{\dagger}(-{\bf k}).
\end{equation}
Spinon states created by $\zeta_{\sigma}^{\dagger}({\bf k})$ have
``bare'' energies $\epsilon^{\zeta} ({\bf k})$, having a V-shape zero
minimum at ${\bf k}={\bf k}_0$. Bose condensation results in an
antiferromagnetic (AF) order of wave vector ${\bf Q}=2{\bf k}_0 = ({\pi
\over a} , {\pi \over a})$. Within the lattice Brillouin zone (BZ) there
are four inequivalent possibilities for ${\bf k}_0$: $\pm({\pi \over 2a}
, {\pi \over 2a})$ and $\pm({\pi \over 2a} , -{\pi \over 2a})$, thus
introducing a broken symmetry. One has \cite{Ashk3}: 
\begin{eqnarray}
\cosh{(\xi_{{\bf k}})} &\to& \cases{+\infty \;,& for ${\bf k} \to {\bf
k}_0$,\cr 1 \;,& for ${\bf k}$ far from ${\bf k}_0$,\cr} \nonumber \\
\sinh{(\xi_{{\bf k}})} &\to& \cases{-\cosh{(\xi_{{\bf k}})} \;,& for
${\bf k} \to {\bf k}_0$,\cr 0 \;,& for ${\bf k}$ far from ${\bf
k}_0$.\cr} 
\end{eqnarray} 

The dynamic stripe-like inhomogeneities are approached adiabatically,
treating them statically with respect to the electrons dynamics. Their
underlying structure is characterized \cite{Tran1} by narrow charged
stripes forming antiphase domain walls between wider AF stripes. Within
the one-dimensional charged stripes it is justified to use the
spin-charge separation approximation under which two-particle
spinon-holon (spinon-excession) Green's functions are decoupled into
single-auxiliary-particle Green's functions. Holons (excessions) within
the charged stripes are referred to as ``stripons'' (of charge $-{\rm
e}$), created by $p^{\dagger}_{\mu}({\bf k})$. 

Localized stripon states are assumed at the zeroth order, due to the
disordered one-dimensional nature of the charged stripes. Their ${\bf
k}$ wave vectors present ${\bf k}$-symmetrized combinations of localized
states to be treated in a perturbation expansion when coupling to the
other fields is considered. 

Away from the charged stripes, creation operators of approximate fermion
basis states of coupled holon-spinon and excession-spinon pairs are
constructed \cite{Ashk1}. Together with the small-$U$ states they form,
within the auxiliary space, a basis to ``quasi-electron'' (QE) states,
created by $q_{\iota\sigma}^{\dagger}({\bf k})$. The bare QE energies
$\epsilon^q_{\iota} ({\bf k})$ form quasi-continuous ranges of bands
within the BZ. 

When the cuprates are doped, such QE states are transferred from the
upper and lower Hubbard bands to the vicinity of the Fermi level
($E_{_{\rm F}}$). The amount of states transferred is increasing with
the doping level, moving from the insulating to the metallic side of the
Mott transition regime. 

Hopping and hybridization terms introduce strong coupling between the
QE, stripon and spinon fields, which is expressed by a coupling
Hamiltonian of the form (for p-type cuprates): 
\begin{eqnarray}
{\cal H}^{\prime} &=& {1 \over \sqrt{N}} \sum_{\iota\lambda\mu\sigma}
\sum_{{\bf k}, {\bf k}^{\prime}} \big\{\sigma
\epsilon^{qp}_{\iota\lambda\mu}(\sigma{\bf k}, \sigma{\bf k}^{\prime})
q_{\iota\sigma}^{\dagger}({\bf k}) p_{\mu}^{\dagg}({\bf k}^{\prime})
[\cosh{(\xi_{\lambda,\sigma({\bf k} - {\bf k}^{\prime})})} \nonumber \\
&\ &\times \zeta_{\lambda\sigma}^{\dagg}({\bf k} - {\bf k}^{\prime}) +
\sinh{(\xi_{\lambda,\sigma({\bf k} - {\bf k}^{\prime})})}
\zeta_{\lambda,-\sigma}^{\dagger}({\bf k}^{\prime} - {\bf k})] + h.c.
\big\}. 
\end{eqnarray} 
Using the formalism of Green's functions (${\cal G}$), the QE, stripon
and spinon propagators are couples by a vertex introduced through ${\cal
H}^{\prime}$ \cite{Ashk4}. 

The stripe-like inhomogeneities are strongly coupled to the lattice, and
it is the presence of stripons which creates the charged stripes within
them. Consequently, processes involving transitions between stripon and
QE states (which, through ${\cal H}^{\prime}$, are followed by the
emission and/or absorption of spinons) involve also lattice
displacements. Since the hopping and hybridization terms depend on the
atomic positions \cite{Andersen}, the effect of these lattice
displacements can be expressed by modifying ${\cal H}^{\prime}$ in
Eq.~(3), adding to each of the spinon creation and annihilation
operators there a term in which the spinon operator is multiplied by a
lattice displacement operator. This introduces a new vertex due to which
the spinons are renormalized, becoming ``dressed'' by phonons, and thus
carry some lattice distortion in addition to spin. Such phonon-dressed
spinons are referred to as ``svivons'', and the physical effect of the
new vertex, which is most relevant here, is assumed to be the
replacement the spinons by svivons in the ${\cal H}^{\prime}$ vertex. 

\begin{figure}[t] 
\begin{center}
\includegraphics[width=3.25in]{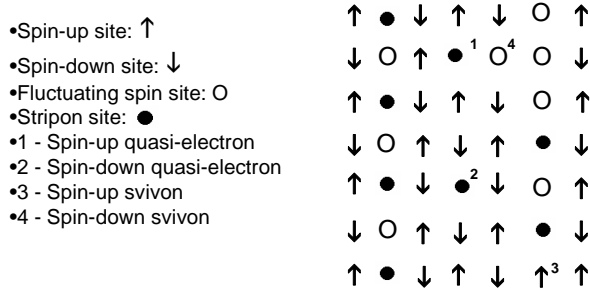}
\end{center}
\caption{An adiabatic ``snapshot'' of a stripe-like inhomogeneity and 
carriers within a CuO$_2$ plane.} 
\label{fig1}
\end{figure}

The physical signature of the auxiliary fields, within the
$t$--$t^{\prime}$--$J$ model, is demonstrated in Fig.~1, where an
adiabatic ``snapshot'' of a section of a CuO$_2$ plane, including a
stripe-like inhomogeneity, is shown. Within the adiabatic time scale a
site is ``spinless'' either if it is ``charged'', removing the spinned
electron/hole on it (as in ``stripon sites'' in Fig.~1), or if the spin
is fluctuating on a shorter time scale (due to, {\it e.g.}, being in a
singlet spin pair). In this description, a site stripon excitation
represents a transition between these two types of a spinless site
within the charged stripes, a site svivon excitation represents a
transition between a spinned site and a fluctuating-spin spinless site,
and a site QE excitation represents a transition between a spinned site
and a charged spinless site within the AF stripes. The dynamics of these
sites is ignored in Fig.~1 for demonstration purposes. 

\section{Auxiliary Spectral Functions}

Self-consistent expressions are derived \cite{Ashk1} for the spectral
functions $A^q$, $A^p$, and $A^{\zeta}$ for the QE's, stripons, and
svivons, respectively [$A(\omega) \equiv \Im {\cal G}(\omega-i0^+) /
\pi$], including the scattering rates $\Gamma^q$, $\Gamma^p$, and
$\Gamma^{\zeta}$ [$\Gamma(\omega) \equiv 2\Im \Sigma(\omega-i0^+)$], and
the real parts of the self-energies $\Re\Sigma^q$, $\Re\Sigma^p$ and
$\Re\Sigma^{\zeta}$. Since the stripon bandwidth turns out to be
considerably smaller than the QE and svivon bandwidths, a phase-space
argument could be used, as in the Migdal theorem, to ignore vertex
corrections to the ${\cal H}^{\prime}$ vertex. 

The expressions are derived for the intermediary energy range (and the
$T\to 0$ limit), where the high energy range ($\gta 0.5\;$eV) is treated
by introducing cut-off integration limits at $\pm\omega_c$ (resulting in
spurious logarithmic divergencies at $\pm\omega_c$), and the low energy
range ($\lta 0.02\;$eV) introduces non-analytic behavior at
``zero-energy''. The ${\bf k}$ dependence is omitted in the expressions
for simplicity, and all the coefficients appearing in them are positive.
The expressions for the auxiliary spectral functions are: 
\begin{eqnarray}
A^q(\omega) &\cong& \cases{a^q_+\omega + b^q_+ \;,& for $\omega>0$,\cr
-a^q_-\omega + b^q_- \;,& for $\omega<0$,\cr} \\ 
A^p(\omega) &\cong& \delta(\omega), \\ 
A^{\zeta}(\omega) &\cong& \cases{a^{\zeta}_+\omega + b^{\zeta}_+ \;,&
for $\omega>0$,\cr a^{\zeta}_-\omega - b^{\zeta}_- \;,& for
$\omega<0$.\cr} 
\end{eqnarray} 
Analyticity is restored in the low-energy range, and specifically
$A^{\zeta}(\omega=0) = 0$. Special behavior occurs for svivons around
${\bf k}_0$. The expressions for the QE and svivon scattering rates are:
\begin{eqnarray}
{\Gamma^q(\omega) \over 2\pi} &\cong& \cases{c^q_+\omega + d^q_+ \;,&
for $\omega>0$,\cr -c^q_-\omega + d^q_- \;,& for $\omega<0$,\cr} \\ 
{\Gamma^{\zeta}(\omega) \over 2\pi} &\cong& \cases{c^{\zeta}_+\omega +
d^{\zeta}_+ \;,& for $\omega>0$,\cr c^{\zeta}_-\omega - d^{\zeta}_- \;,&
for $\omega<0$,\cr} 
\end{eqnarray} 
and those for the real parts of their self energies are:
\begin{eqnarray}
-\Re\Sigma^q(\omega) &\cong& \omega_c(c^q_+ - c^q_-) + \big(d^q_+
\ln{\big|{\omega - \omega_c \over \omega}\big|} - d^q_- \ln{\big|{\omega
+ \omega_c \over \omega}\big|}\big) \nonumber \\ &\ &+ \omega\big(c^q_+
\ln{\big|{\omega - \omega_c \over \omega}\big|} + c^q_- \ln{\big|{\omega
+ \omega_c \over \omega}\big|}\big), \\ 
-\Re\Sigma^{\zeta}(\omega) &\cong& \omega_c(c^{\zeta}_+ + c^{\zeta}_-) +
\big(d^{\zeta}_+ \ln{\big|{\omega - \omega_c \over \omega}\big|} +
d^{\zeta}_- \ln{\big|{\omega + \omega_c \over \omega}\big|}\big)
\nonumber \\ &\ &+ \omega\big(c^{\zeta}_+ \ln{\big|{\omega - \omega_c
\over \omega}\big|} - c^{\zeta}_- \ln{\big|{\omega + \omega_c \over
\omega}\big|}\big). 
\end{eqnarray} 
The logarithmic divergencies at $\omega=0$ are truncated by analyticity
in the low-energy range. 

For the case of p-type cuprates the following inequalities exist between
the coefficients: 
\begin{eqnarray}
a^q_+>a^q_-,\ \ \ \ \  b^q_+>b^q_-,\ \ \ \ \   c^q_+>c^q_-,\ \ \ \ \
d^q_+>d^q_-,&\ & \\ a^{\zeta}_+>a^{\zeta}_-,\ \ \ \ \
b^{\zeta}_+>b^{\zeta}_-,\ \ \ \ \ c^{\zeta}_+>c^{\zeta}_-,\ \ \ \ \
d^{\zeta}_+>d^{\zeta}_-.&\ & 
\end{eqnarray} 
For ``real'' n-type cuprates, in which the stripons are based on
excession and not holon states, the direction of the inequalities is
reversed for the QE coefficients (11), but stays the same for the svivon
coefficients (12). Deviations from these inequalities, especially for
the $a_{\pm}$ and $b_{\pm}$ coefficients, could occur due to
band-structure effects, and at specific ${\bf k}$ points; by Eq.~(2)
they almost disappear for svivons close to point ${\bf k}_0$. 

The auxiliary-particle energies are renormalized (from $\epsilon$ to
$\bar\epsilon$) through: $\bar\epsilon = \epsilon +
\Re\Sigma(\bar\epsilon)$. This renormalization is particularly strong
for the stripon energies, and their bandwidth drops down to the low
energy range [thus having a $\delta$-function for $A^p(\omega)$ in
Eq.~(5)]. 

\begin{figure}[t] 
\begin{center}
\includegraphics[width=3.25in]{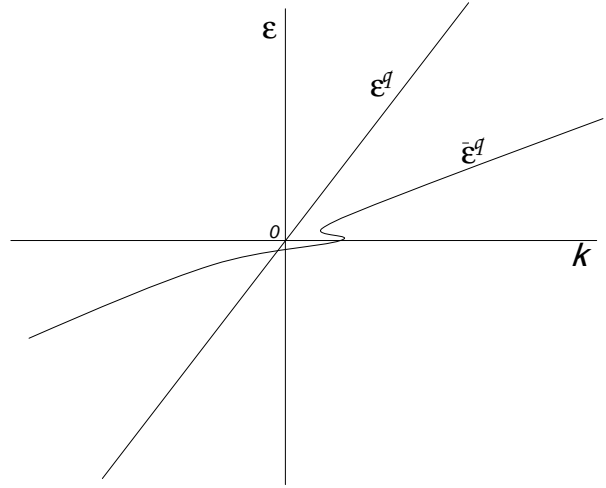}
\end{center}
\caption{A typical self-energy renormalization of the QE energies, for
p-type cuprates.} 
\label{fig2}
\end{figure}

A typical renormalization of the QE energies (for p-type cuprates),
around zero energy, is shown in Fig.~2. The kink-like behavior around
zero energy is a consequence of the logarithmic singularity at
$\omega=0$ (truncated in the low-energy range) in the $(d^q_+ -
d^q_-)\ln{|\omega|}$ term in Eq.~(9) for $\Re\Sigma^q$. The asymmetry
between positive and negative energies is a consequence of inequality
(11), and this asymmetry is expected to be inverted for real n-type
cuprates. 

\begin{figure}[t] 
\begin{center}
\includegraphics[width=3.25in]{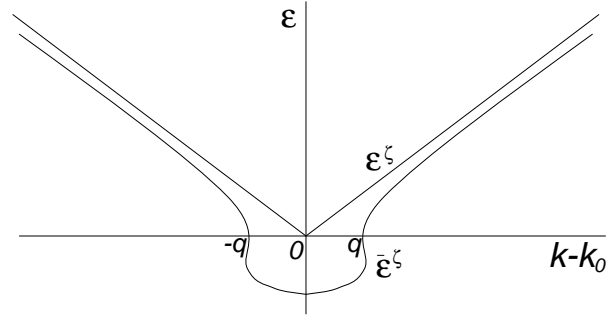}
\end{center}
\caption{A typical self-energy renormalization of the svivon energies
around the minimum at ${\bf k}_0$.} 
\label{fig3}
\end{figure}

A typical renormalization of the svivon energies, around the V-shape
zero minimum of $\epsilon^{\zeta}$ at ${\bf k}_0$, is shown in Fig.~3.
The significant effect results from the $(d^{\zeta}_+ +
d^{\zeta}_-)\ln{|\omega|}$ term in Eq.~(10) for
$\Re\Sigma^{\zeta}(\omega)$, contributing a logarithmic singularity at
$\omega=0$ (which is truncated in the low-energy range). By Eq.~(8)
$\bar\epsilon^{\zeta}$ is expected to have a considerable linewidth
around ${\bf k}_0$ (except where it crosses zero). But the existence of
a pairing gap in the low-energy QE and stripon states, coupled by
svivons around ${\bf k}_0$, disables the scattering processes causing
the linewidth near the negative minimum of $\bar\epsilon^{\zeta}$ at
${\bf k}_0$. 

This renormalization of the svivon energies changes the physical
signature of their Bose condensation from an AF order to the observed
stripe-like inhomogeneities (including both their spin and lattice
aspects). The structure of $\bar\epsilon^{\zeta}$ around its minimum at
${\bf k}_0$ determines the structure of these inhomogeneities, and their
dynamics depends on the linewidth of $\bar\epsilon^{\zeta}$
around ${\bf k}_0$, becoming slower (and thus detectable) in a pairing
state. 

As was mentioned above, the stripon states are based on localized states
in the charged stripe-like inhomogeneities (see Fig.~1), to which some
itineracy is introduced by coupling to the QE and svivon states through
${\cal H}^{\prime}$. This implies that their translational symmetry is
lower than that of the basic lattice, resulting in a mixture of ${\bf
k}$ values of the lattice BZ. Since the charged stripes (where the bare
stripon states reside) occupy about a quarter of the CuO$_2$ plane, the
number of stripon states should be about a quarter of the number of
states in this BZ. 

The BZ ${\bf k}$ values mostly contributing to the stripon states
reflect the structural nature of the stripe-like inhomogeneities, on one
hand, and the minimization of free energy due to the ${\cal
H}^{\prime}$-coupling, on the other hand. Such a minimization is
achieved when the stripon ${\bf k}$ values are mainly at BZ areas where
the energetic effect of their coupling is the strongest. This would
occur for optimal stripon coupling with svivons around ${\bf k}_0$ (see
the behavior of $\bar\epsilon^{\zeta}$ there in Fig.~3), and with QE's
at BZ areas of highest density of states (DOS) close to $E_{_{\rm F}}$,
which are found in most of the cuprates around the ``antinodal'' points
$({\pi \over a} , 0)$ and $(0 , {\pi \over a})$. 

If (from its four possibilities) ${\bf k}_0$ were chosen at $({\pi \over
2a} , {\pi \over 2a})$, then in order to minimize free energy, the BZ
areas (in most of the cuprates) which the ${\bf k}$ values contributing
to the stripon states should mostly come from, would be at about a
quarter of the BZ around $\pm{\bf k}^p = \pm({\pi \over 2a} , -{\pi
\over 2a})$ (thus the antinodal points are at ${\bf k}^p \pm {\bf
k}_0$). Stripon states corresponding to an equal mixture of ${\bf k}^p$
and $-{\bf k}^p$ states are created by: 
\begin{eqnarray}
p^{\dagger}_{\rm e}(\pm{\bf k}^p) &\propto& \sum_i
p^{\dagger}_i\cos{\Big[{\pi \over 2a}(x_i-y_i)\Big]}, \nonumber \\
p^{\dagger}_{\rm o}(\pm{\bf k}^p) &\propto& \sum_i
p^{\dagger}_i\sin{\Big[{\pi \over 2a}(x_i-y_i)\Big]}. 
\end{eqnarray} 
In the stripe-like inhomogeneity shown in Fig.1 (where the stripes are
directed in the $y$-direction), one has $(x_i,y_i)=(4am,an)$, where $m$
and $n$ are integers, and thus $p^{\dagger}_{\rm e}(\pm{\bf k}^p)$
creates a state with non-zero amplitudes in the even $n$ points, while
$p^{\dagger}_{\rm o}(\pm{\bf k}^p)$ creates a state with non-zero
amplitudes in the odd $n$ points. A similar result would be obtained
also in lattice areas where the stripes are directed in the
$x$-direction. It will be shown below that pairing occurs between
stripon states close to the ones in Eq.~(13). 

\section{Electron Spectrum}

Spectroscopic measurements (as in ARPES) based  on the transfer of
electrons into, or out of, the crystal, are determined by the electron's
spectral function $A_e$. Projecting the spectral functions from the
auxiliary to the physical space \cite{Ashk1}, $A_e$ is expressed in
terms of QE ($A^q$) and convoluted stripon-svivon ($A^p A^{\zeta}$)
terms. From the quasi-continuum of QE bands, only few, closely related
to those of physical electron, contribute ``coherent'' bands, while the
others contribute an ``incoherent'' background to $A_e$. 

The electron bands are specified by ${\bf k}$ vectors within the lattice
BZ, though the stripe-like inhomogeneities introduce a perturbation of
lower periodicity, reflected {\it e.g.} in ``shadow bands''. These bands
(as well as the incoherent background in $A_e$) include hybridized
contributions of QE states and convoluted stripon--svivon states. As in
Eq.~(7) for $\Gamma^q$, the electronic bandwidths have a $\propto\omega$
and a constant term, in agreement with experiment. 

A significant stripon--svivon contribution to $A_e$ close to $E_{_{\rm
F}}$ (at energies around $\bar\epsilon^p \pm \bar\epsilon^{\zeta}$) is
obtained with svivons around their energy minimum at ${\bf k}_0$ (see
Fig.~3). As was discussed above, such a contribution should be found (in
most cuprates) in BZ areas around the antinodal points, as has been
widely observed (see {\it e.g.} Ref.~\cite{Yoshida1}). 

Such type of a stripon--svivon contribution is not expected close to
``nodal'' Fermi surface (FS) crossing points, in the vicinity of
$\pm({\pi \over 2a} , \pm{\pi \over 2a})$. Thus the behavior of the
electron bands there should be similar to that of the QE bands
$\bar\epsilon^q$ (see Fig.~2), having a kink closely below $E_{_{\rm
F}}$, as has been observed in ARPES \cite{Lanzara1,Johnson}. The
observed kink has been attributed to the coupling of electrons to
phonons \cite{Lanzara1} or to the neutron scattering resonance mode
\cite{Johnson}. However, such a coupling would generally result in two
opposite changes in the band slope (below and above the coupled
excitation energy) below $E_{_{\rm F}}$, while the experimental kink
looks more consistent with one change in slope below $E_{_{\rm F}}$, as
in Fig.~2. 

This kink was not found in measurements in the n-type cuprate NCCO
\cite{Armitage1}, which is consistent with the prediction here (suggested
by the author earlier \cite{Ashk2}) that in real n-type cuprates this
kink should be above, and not below $E_{_{\rm F}}$ (where ARPES
measurements are relevant). Also, there appears to be a sharp upturn in
the ARPES band in NCCO \cite{Armitage1} very close to $E_{_{\rm F}}$
(believed there to be an artifact), which is expected here as the kink
is approached from the other side of $E_{_{\rm F}}$ (see Fig.~2). 

Measurements of the doping-dependence of the slopes of the electron
bands around the nodal points \cite{Zhou1} show almost no change with
doping of the slope very close to $E_{_{\rm F}}$ from below (thus
including the kink effect). Here this low-energy slope depends on the
nature of the low-energy truncation of the logarithmic singularity in
Eq. (9); thus it depends of the width $\omega^p$ of the stripon band.
Our analysis for the thermoelectric power \cite{Ashk2}, discussed below,
predicts $\omega^p \sim 0.02\;$eV, with very weak decrease with doping,
which is consistent with the observed slope behavior. 

The nodal kink, discussed above, shows almost no change when the
temperature is lowered below $T_c$, as is expected here. A different
type of a kink has been observed around the antinodal points
\cite{Gromko,Sato}, showing a very strong temperature dependence, where
its major part appears only below $T_c$. Within the present analysis,
the temperature dependent part of this antinodal kink originates from
the stripon--svivon contribution to the electron bands (discussed
above), while the temperature-independent part is of the same origin as
the nodal kink (thus due to the QE contribution to the electron bands). 

As will be discussed below, the opening of a superconducting (SC) gap
causes a decrease in the svivon linewidth around the energy minimum at
${\bf k}_0$, resulting in the narrowing of the stripon--svivon
contribution to the antinodal electron bands below $T_c$. This is
expressed by the appearance of the temperature-dependent antinodal kink,
as well as a peak-dip-hump structure, below $T_c$ (see discussion
below). The appearance of this antinodal kink away from the nodal point
on the FS is viewed in ARPES studies, above and below $T_c$
\cite{Lanzara2}, of FS crossings between the nodal point and half way
towards the antinodal point. 

The existence of high electron DOS close to $E_{_{\rm F}}$ around the
antinodal points is actually a consequence of the hybridization between
QE's and convoluted stripon--svivon states of svivons around ${\bf
k}_0$. This has been viewed in ARPES \cite{Zhou2} as ``an extra low 
energy scattering mechanism'' around the antinodal points. As was
discussed above, this could occur when the stripons reside around points
$\pm{\bf k}^p$, which by Eq. (13) is consistent with ``vertical stripes''
(as those shown in Fig.~1). And indeed, in ARPES measurements in LSCO
\cite{Yoshida2}, including both high and low (non-SC) doping levels
(which are characterized by ``diagonal stripes''), a contribution of
$A_e$ close to $E_{_{\rm F}}$ is observed around the nodal points for
all doping levels, while around the antinodal points it is observed only
for doping levels where the stripes are vertical. 

As was demonstrated in Ref.~\cite{Ashk1}, $t^{\prime}$ hopping processes
(between next-nearest-neighbor sites -- along the diagonal) in a CuO$_2$
plane, can take place without disrupting the AF order, and thus are only
weakly affected by stripes and stripons, while in order for $t$ hopping
processes (between nearest-neighbor sites) to occur without the
disruption of the AF order, it is essential for stripons in vertical
stripes of $4a$ separation (as in Fig.~1) to be involved. The existence
of nodal electron states close to $E_{_{\rm F}}$ is mainly due to
$t^{\prime}$ processes, while the existence of antinodal electron states
close to $E_{_{\rm F}}$ is mainly due to $t$ processes. 

\section{The Neutron Resonance Mode}

The imaginary part of the spin susceptibility $\chi^{\prime\prime}({\bf
q}, \omega)$ (at wave vector ${\bf q}$ and energy $\omega$) has a major
contribution from double-svivon excitations which can be expressed
\cite{Ashk1} as: 
\begin{eqnarray}
\chi^{\prime\prime}({\bf q}, \omega) &\sim& \sum_{{\bf k}}
\sinh{(2\xi_{{\bf k}})} \sinh{(2\xi_{{\bf q} - {\bf k}})} \int
d\omega^{\prime} A^{\zeta}({\bf k}, \omega^{\prime}) \nonumber \\ &\ &
\times \big\{ A^{\zeta}({\bf q} - {\bf k}, -\omega -\omega^{\prime}) -
A^{\zeta}({\bf q} - {\bf k}, \omega - \omega^{\prime}) \nonumber \\ &\ &
+ 2A^{\zeta}({\bf q} - {\bf k}, \omega^{\prime} - \omega)
[b_{_T}(\omega^{\prime} - \omega) - b_{_T}(\omega^{\prime}] \big\}. 
\end{eqnarray}
Due to the $\sinh{(2\xi)}$ factors, and Eq.~(2), large contributions to
$\chi^{\prime\prime}$ are obtained when both ${\bf k}$ and ${\bf q} -
{\bf k}$ are close to ${\bf k}_0$. The effect of the negative minimum of
$\bar\epsilon^{\zeta}({\bf k})$ at ${\bf k}_0$, (see Fig.~3), and
specifically in the SC state, where its linewidth is often small, is the
existence of a peak in $\chi^{\prime\prime}({\bf q}, \omega)$ at ${\bf
q} = 2{\bf k}_0 = {\bf Q}$ (the AF wave vector) and $\omega =
-2\bar\epsilon^{\zeta}({\bf k}_0)$. This peak is consistent with the
neutron resonance mode [of energy $E_{\rm res} =
-2\bar\epsilon^{\zeta}({\bf k}_0)$], often found in the high-$T_c$
cuprate at $\sim 0.04\;$eV \cite{Bourges1,Reznik}. It is expected here
that the energy of this resonance mode has a local maximum, as a
function of ${\bf k}$, at ${\bf k}={\bf Q}$ [since
$\bar\epsilon^{\zeta}$ has a minimum at ${\bf k}_0$]; however, also a
branch of the mode with energy rising with ${\bf k} - {\bf Q}$ is
expected due to the range where $\bar\epsilon^{\zeta}({\bf k})$ is
positive and rising. And indeed, measurements in YBCO
\cite{Bourges1,Reznik} show a neutron-scattering peak branch dispersing
downward (from the ${\bf k}={\bf Q}$ value), and also one dispersing
upward \cite{Reznik}. An approximate circular symmetry around ${\bf
k}={\bf Q}$ is obtained \cite{Reznik}, as is expected here. 

The incommensurate low-energy neutron-scattering peaks, corresponding to
the stripe-like inhomogeneities \cite{Tran1}, occur at points ${\bf Q}
\pm 2{\bf q}$, where the slope of the low-energy
$\bar\epsilon^{\zeta}({\bf k})$ is not too steep (see Fig.~3). In the
LSCO system the resonance energy at ${\bf k}={\bf Q}$ is higher than the
maximal SC gap, and thus a sharp peak is not observed there, being too
wide (see discussion below); however, sharp peaks have been observed
\cite{Tran2,Wakimoto,Christ} at incommensurate ${\bf Q} \pm 2{\bf q}$
points, where the energy is lower than the maximal SC gap. In the LBCO
system, the energy dependence of the peak (whether it is sharp or wide)
with ${\bf q}$ was also found \cite{Tran3} to have branches dispersing
both downward and upward around ${\bf k}={\bf Q}$. A similar behavior
was observed in YBCO$_{6.6}$ \cite{Hayden}. 

Thus, both the resonance mode at ${\bf k}={\bf Q}$, and the excitations
at incommensurate ${\bf Q} \pm 2{\bf q}$, are double-svivon excitations,
around its energy minimum at ${\bf k}_0$, shown in Fig.~3. These are
excitations towards the destruction of the stripe-like inhomogeneities;
their width determines the speed of the inhomogeneities dynamics, and
they exist for stoichiometries where SC exists \cite{Wakimoto}. Because
of the lattice dressing of svivons, these double-svivon excitations are
expected to be lattice-dressed spin excitations, and indeed, Cu--O
optical phonon modes have been found to be involved in such excitations
\cite{Egami}. 

Spin excitations in bilayer cuprates are expected \cite{Ashk1} to have
either odd or even symmetry, with respect to the layers exchange. For
odd symmetry one gets results similar to those obtained in the
single-layer approach discussed above, while for even symmetry one gets
a mode whose energy has a minimum at ${\bf k} = {\bf Q}$, in agreement
with experiment \cite{Bourges2}. 

\section{Transport Properties}

Transport properties, unlike {\it e.g.} ARPES, measure the electrons
{\it within} the crystal, and thus can detect the small energy scale of
the stripons, without convoluting them with svivons. 

Normal-state transport expressions, {\it not} including the effect of
the pseudogap (PG), were derived \cite{Ashk2} using linear response
theory, where the zero-energy singularities in the auxiliary spectral
functions in Eqs. (4--6) are smoothened in the low-energy range, through
Taylor expansion [imposing $A^{\zeta}(\omega=0)=0$]. In this derivation
it was taken into account that the electric current can be expressed as
a sum ${\bf j} = {\bf j}^q_0 + {\bf j}^p_0$ of contributions of bare QE
and stripon states, respectively, and that ${\bf j}^p_0 \cong 0$ since
the bare stripon states are localized. Thus the contributions to the
current of both the QE and the stripon dressed (thus coupled) states
originate from ${\bf j}^q_0$. 

Results for the electrical resistivity $\rho$, the Hall constant
$R_{_{\rm H}}$, the Hall number $n_{_{\rm H}} = 1/{\rm e} R_{_{\rm H}}$,
the Hall angle $\theta_{_{\rm H}}$ (through $\cot{\theta_{_{\rm H}}} =
\rho / R_{_{\rm H}}$), and the thermoelectric power (TEP) $S$, are
presented in Fig.~4. Their anomalous temperature dependencies result
both from the low-energy-range stripon band [Eq.~(5)], modeled by a
``rectangular" shape $A^p$ of width $\omega^p$ and fractional occupancy
$n^p$, and from those of the scattering rates $\Gamma^q(T,\omega=0)$ and
$\Gamma^p(T,\omega=0)$ (see Ref.~\cite{Ashk2}), including also impurity
scattering temperature independent terms. 

\begin{figure}[t] 
\begin{center}
\includegraphics[width=3.25in]{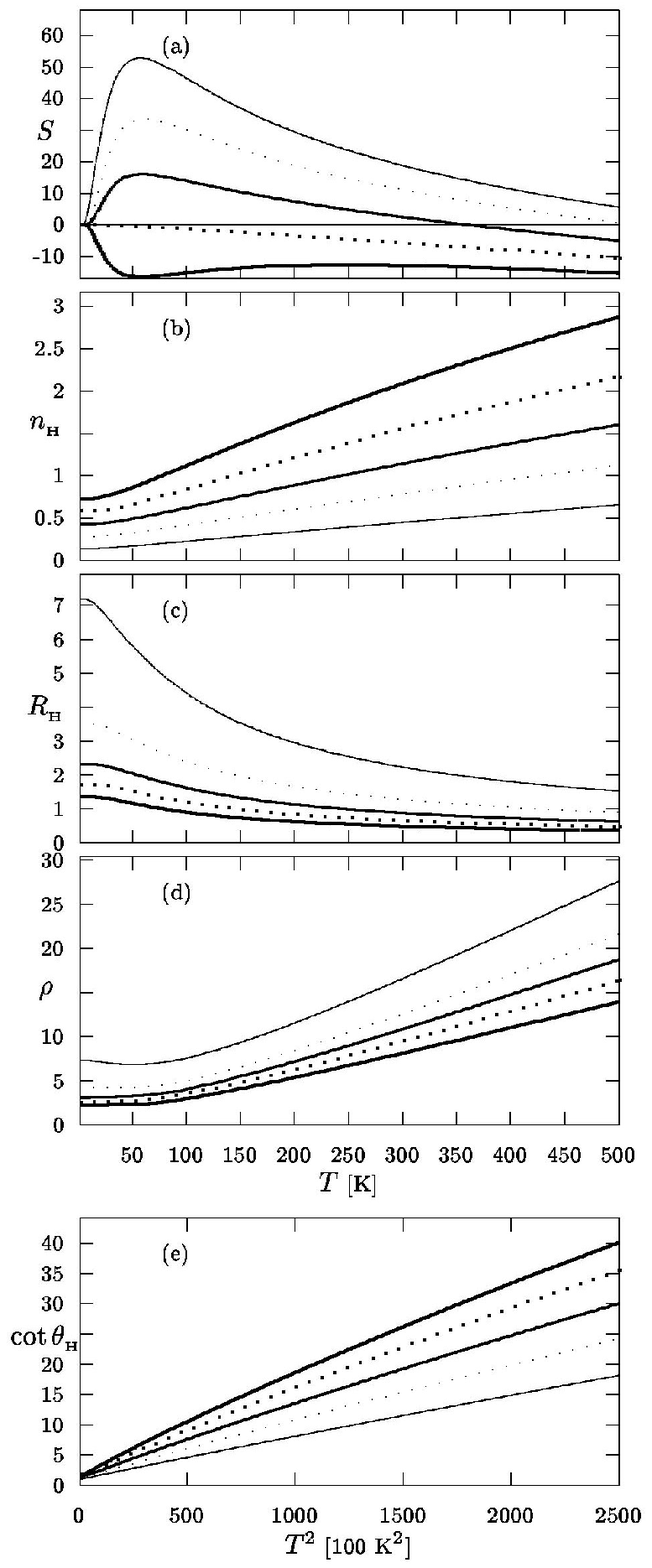}
\end{center}
\caption{The transport coefficients, in arbitrary units [and $\mu$V/K
units for $S$ (a)], for: $n^p$=0.8,0.7,0.6,0.5,0.4;
$10000N^q_e$=20,23,26,29,32; $\omega^p$[K]=200,190,180,170,160;
$n^p_{_{\rm H}}$=0.1,0.2,0.3,0.4,0.5; $n^q_{_{\rm H}}$=6,7,8,9,10;
$S^q_1$=$-$0.025; $\gamma^p_0$=500; $\gamma^p_2$=0.03; $\gamma^q_0$=5;
$\gamma^q_1$=0.2. The last values correspond to the thickest lines.} 
\label{fig4}
\end{figure}

The transport results in Fig.~4 correspond to five stoichiometries of
p-type cuprates, ranging from $n^p=0.8$, corresponding to the underdoped
(UD) regime, to $n^p=0.4$, corresponding to the overdoped (OD) regime.
The parameter $N^q_e$ corresponds to the QE contribution to the
electrons DOS at $E_{_{\rm F}}$. It is assumed to increase with the
doping level $x$, reflecting transfer of QE spectral weight from the
upper and lower Hubbard bands towards $E_{_{\rm F}}$, while moving from
the insulating to the metallic side of the Mott transition regime.
Consequently $\omega^p$ is assumed to decrease somewhat with doping, due
to stronger renormalization of the stripon energies \cite{Ashk2}. 
 
The parameters $n^q_{_{\rm H}}$ and $n^p_{_{\rm H}}$ represent effective
QE and stripon contributions to the density of charge carriers
(reflected in the Hall number). Since they both contribute through the
current ${\bf j}^q_0$ of the bare QE states, they are expected to have
same sign (corresponding to these states). The values of these
parameters are assumed to increase with $x$ (for the same reason that
$N^q_e$ does). Since the coupling between QE's and stripons grows with
$x$, and the increase of $N^q_e$ with $x$ is slower than that of
$1-n^p$, it is assumed that the increase of $n^p_{_{\rm H}}$ with $x$ is
faster than that of $1-n^p$, which is faster than that of $n^q_{_{\rm
H}}$. 

Doping-independent values are assumed for the QE TEP parameter $S^q_1$
\cite{Ashk2} [which is normally negative for p-type cuprates by the
inequality (11)], and for the stripon and QE scattering rate parameters
$\gamma^p_0$, $\gamma^p_2$, $\gamma^q_0$, and $\gamma^q_1$ \cite{Ashk2}.

The TEP results depend strongly on $n^p$, and reproduce very well the
doping-dependent experimental behavior \cite{Fisher,Tanaka}. The
position of the maximum in $S$ depends on the choice of $\omega^p$, and
it may occur below or above $T_c$ (the existence of a PG may shift it to
a higher temperature than predicted here). 

Also the results for the Hall coefficients in Fig.~4 reproduce very well
the experimental behavior \cite{Kubo,Hwang}. The anomalous temperature
dependence of the $n_{_{\rm H}}$ is due to the growing role of
$n^q_{_{\rm H}}$ in its determination with increasing $T$, being
dominantly determined by $n^p_{_{\rm H}}$ at $T=0$. 

The temperature dependence of the resistivity in Fig.~4(d) is linear at
high $T$, becoming ``sublinear'' at low $T$ (for all stoichiometries),
while experimentally \cite{Takagi} the low-$T$ behavior crosses over
from ``superlinarity'' in the UD regime, to sublinearity in the OD
regime (being linear at low $T$ for optimally doped cuprates). The
superlinear behavior is being generally understood as the effect of the
PG (not considered here), and the crossover to sublinear behavior
(predicted here) in the OD regime is a natural consequence of the
disappearance of the PG with increasing $x$. 

The TEP in real n-type cuprates is normally expected \cite{Ashk2} to
behave similarly to the TEP in p-type cuprates, but with an opposite
sign and slope. Results for NCCO \cite{Takeda} show such behavior for
low doping levels, but in SC doping levels the slope of $S$ changes from
positive to negative, and its behavior resembles that of OD p-type
cuprates, shown in Fig.~4(a). This led \cite{Ashk2} to the suggestion
that NCCO may be not a real n-type cuprate, its stripons being based on
holon states (like in p-type cuprates). More recent measurements on the
n-type infinite-CuO$_2$-layer SLCO \cite{Williams} do show TEP results
for an SC cuprate which have the opposite sign and slope than those of
Fig.~4(a) (for p-type cuprates), as is expected for real n-type cuprates
(thus with stripons based on excession states). 

As was discussed above, the absence of a kink in the nodal band below
$E_{_{\rm F}}$ \cite{Armitage1} in NCCO, supports the possibility that
it is also a real n-type cuprate. It is possible that the change in the
sign of the TEP slope in NCCO with doping is an anomalous band-structure
effect, probably associated with the peculiar evolution of its FS with
doping, detected in ARPES \cite{Armitage2}. The position of the kink
(below or above $E_{_{\rm F}}$) is determined by the inequality (11)
between $d^q_+$ and $d^q_-$, which is less susceptible to band-structure
effects than the inequality (11) between $b^q_+$ and $b^q_-$, determining
the sign of the TEP slope. Anomalous behavior is observed also in the
Hall constant of NCCO \cite{Takeda}, which changes its sign with
temperature in the stoichiometries where the sign of the slope of the
TEP has changed. 

\section{Hopping-Induced Pairing}

As was demonstrated in Ref.~\cite{Ashk1}, the ${\cal H}^{\prime}$ vertex
enables inter-stripe stripon hopping, through intermediary QE--svivon
states. This vertex was also demonstrated \cite{Ashk1} to enable
inter-stripe hopping of pairs of neighboring stripons through
intermediary states of pairs of opposite-spin QE's, obtained by the
exchange of svivons. The pair hopping was shown to  result in a gain in
inter-stripe hopping energy (compared to the hopping of two uncorrelated
stripons), avoiding intermediary svivon excitations. Furthermore, the
hybridization of the QE's with orbitals, beyond the
$t$--$t^{\prime}$--$J$ model, results in further gain in both
intra-plane and inter-plane hopping energy. 

This provides a pairing scheme based on transitions between pair states
of stripons and QE's through the exchange of svivons. The inter-plane
pair hopping, introduced within this scheme, is consistent with $c$-axis
optical conductivity results \cite{marel1}, revealing (in addition to
the opening of the SC gap), the increase of the spectral weight in the
mid-IR range (well above the gap) below $T_c$. This effect has been
observed both in bilayer and single-layer cuprates, and proposed
\cite{marel1} to be the signature of a $c$-oriented collective mode
emerging (or sharpening) below $T_c$. The existence of such a mode below
$T_c$ is due to the hopping of pairs in the $c$-direction, during their
QE-pair stages, while above $T_c$, $c$-axis hopping of stripons (through
intermediary QE--svivon states) is, at the most, limited to adjacent
CuO$_2$ planes. 

The pairing diagram (sketched in Ref.~\cite{Ashk4}) provides
Eliasherg-type equations, of coupled stripon and QE pairing order
parameters. Coherent pairing occurs \cite{Ashk1} between two subsets of
the QE and stripon states. For QE's these subsets are, naturally, of the
spin-up ($\uparrow$) and spin-down ($\downarrow$) QE's. Since the
stripons are spinless, their subsets should be determined according to a
different criterion. 
 
As was illustrated in Fig.~1, for a CuO$_2$ plane within the
$t$--$t^{\prime}$--$J$ model, the $\uparrow$ QE's can reside on
$\downarrow$ sites, and the $\downarrow$ QE's can reside on $\uparrow$
sites of the stripe-like inhomogeneities. In Fig.~5 an adiabatic
snapshot of an extended section of a stripe-like inhomogeneity is shown,
including an expected crossover between stripe segments directed in the
$a$ and the $b$ directions. Denoted are the {\it available} sites for
the $\uparrow$ and $\downarrow$ QE subsets. Since the QE subsets have a
spatial interpretation in the CuO$_2$ planes (within the adiabatic time
scale) it is natural to choose the stripon subsets also on a spatial
basis, in a manner which optimizes the coupled pairing. 

Since the pairing is optimal between neighboring stripons along the
charged stripes \cite{Ashk1}, the stripon pairing subsets are chosen
such that the nearest neighbors of a site corresponding to one subset
are sites corresponding to the other subset. These subsets are denoted
by $\bigtriangleup$ and $\bigtriangledown$, and the sites {\it
available} for them are shown in Fig.~5 too. Note that each of the 
stripon states created by $p^{\dagger}_{\rm e}(\pm{\bf k}^p)$ and
$p^{\dagger}_{\rm o}(\pm{\bf k}^p)$, defined in Eq. (13), belongs to
a different subset. This indicates that combinations of stripon states 
around points $\pm{\bf k}^p$ in the BZ are consistent with this pairing 
scheme. The required degeneracy of the QE and stripon paired subsets is 
restored by stripes dynamics.

\begin{figure}[t] 
\begin{center}
\includegraphics[width=3.25in]{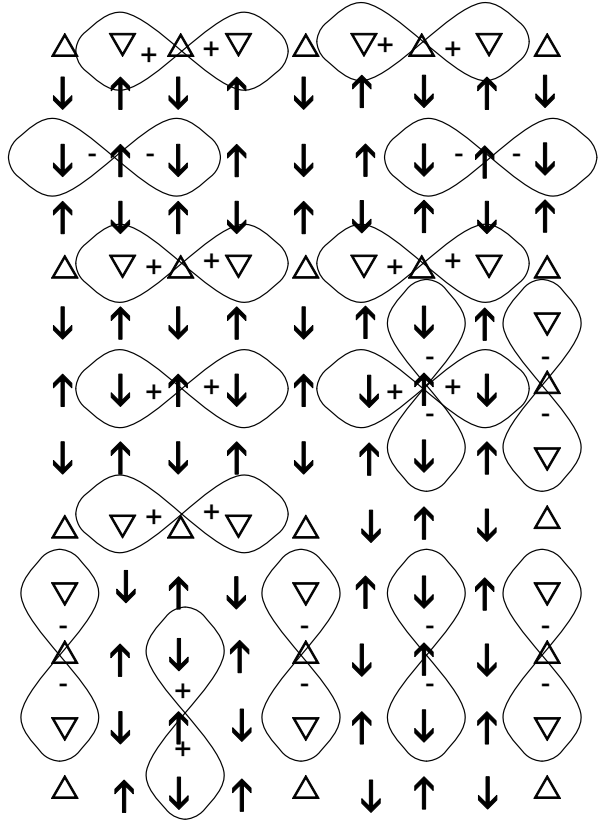}
\end{center}
\caption{An adiabatic snapshot of an extended section of a stripe-like
inhomogeneity, where the available sites for the QE and stripon pairing
subsets are illustrated, as well as sketches demonstrating the local
symmetry of the pairing order parameters $\Phi^q$ and $\Phi^p$.} 
\label{fig5}
\end{figure}

The QE and stripon pair-correlation functions (pairing
order parameters) are defined, within in the position (${\bf r}$)
representation, as: 
\begin{eqnarray}
\Phi^q({\bf r}_1,{\bf r}_2)  &\equiv& \langle q^{\dagg}_{\uparrow}({\bf
r}_1) q^{\dagg}_{\downarrow}({\bf r}_2) \rangle, \\ 
\Phi^p({\bf r}_1,{\bf r}_2)  &\equiv& \langle p^{\dagg}_{\bigtriangleup}
({\bf r}_1) p^{\dagg}_{\bigtriangledown}({\bf r}_2) \rangle. 
\end{eqnarray}
They are coupled to each other through Eliashberg-type equations, which
can be expressed, in the position and the Matzubara ($\omega_n$)
representations, as: 
\begin{eqnarray}
\Phi^q({\bf r}_1,{\bf r}_2,i\omega_n) &=& \sum_{n^{\prime}} \int d{\bf
r}_1^{\prime} \int d{\bf r}_2^{\prime} K^{qp}({\bf r}_1,{\bf r}_2,n;
{\bf r}_1^{\prime},{\bf r}_2^{\prime},n^{\prime}) \nonumber \\ &\
&\times \Phi^p({\bf r}_1^{\prime},{\bf
r}_2^{\prime},i\omega_n^{\prime}), \\ \Phi^p({\bf r}_1,{\bf
r}_2,i\omega_n) &=& \sum_{n^{\prime}} \int d{\bf r}_1^{\prime} \int
d{\bf r}_2^{\prime} K^{pq}({\bf r}_1,{\bf r}_2,n; {\bf
r}_1^{\prime},{\bf r}_2^{\prime},n^{\prime}) \nonumber \\ &\ &\times
\Phi^q({\bf r}_1^{\prime},{\bf r}_2^{\prime},i\omega_n^{\prime}). 
\end{eqnarray} 
Expressions for the kernel functions $K^{qp}$ and $K^{pq}$ are obtained
from the pairing diagrams; they depend on $\Phi^q$ and $\Phi^p$ up to
the temperature where the latter vanish. The combination of Eqs.~(17)
and (18) results in BCS-like equations for both the stripon and the QE
order parameters. The coupling between $\Phi^q$ and $\Phi^p$ results in
maximal pairing between nearest neighbors in the stripe direction, both
for stripons and QE's. The resulting local symmetry of $\Phi^q({\bf
r}_1,{\bf r}_2)$ and $\Phi^p({\bf r}_1,{\bf r}_2)$ (within the adiabatic
time scale) is illustrated in Fig.~5, where point ${\bf r}_1$ is fixed
on selected sites (of the $\uparrow$ and $\bigtriangleup$ QE and stripon
subsets), while point ${\bf r}_2$ is varied over the space including the
nearest neighbors. 

As is illustrated in Fig.~5, the sign of $\Phi^q$ reverses between the
two sides of a charged stripe. This is expected because when two QE
sites on different sides of a charged stripe have a stripon site midway
between them, then one of them is of $\uparrow$ and the other is of
$\downarrow$, and since the exchange of the two fermion operators in the
definition of $\Phi^q$ in Eq.~(15) results in sign reversal, there must
be sign reversal in $\Phi^q$ between the two sites. A similar sign
reversal has been proposed recently by Fine \cite{Fine}. 

The sign of both $\Phi^q$ and $\Phi^p$ is expected to be reversed
between $a$-oriented and $b$-oriented stripe segments meeting in a
``corner'' (shown in Fig.~5). This provides optimal pairing energy,
yielding maximal $|\Phi^q({\bf r}_1,{\bf r}_2)|$ when ${\bf r}_1$ and
${\bf r}_2$ are at nearest neighbor QE sites, and zero $\Phi^q({\bf
r}_1,{\bf r}_2)$ when ${\bf r}_1$ and ${\bf r}_2$ are at next nearest
neighbor sites (where the QE's have the same spin and thus do not pair).
Away from the corner regions, $|\Phi^q({\bf r}_1,{\bf r}_2)|$ is maximal
when ${\bf r}_1$ and ${\bf r}_2$ are at nearest neighbor QE sites along
the stripe direction, but it does not vanish when they are at nearest
neighbor sites perpendicular to the stripe direction (which is not
implied from Fig.~5). 

These symmetry characteristics of $\Phi^q$ and $\Phi^p$ are reflected in
the symmetry of the physical pairing order parameter. The overall
symmetry is expected to be of a $d_{x^2-y^2}$ type; however, the sign
reversal of $\Phi^q$ through the charged stripes, and the lack of
coherence in the details of the dynamic stripe-like inhomogeneities
between different CuO$_2$ planes, is expected to result in features
different from those of a simple $d_{x^2-y^2}$-wave pairing (especially
when the $c$-direction is involved). There is a strong experimental
support in the existence of features of $d_{x^2-y^2}$-wave pairing,
though different features have been reported too. 

\section{Pairing and Coherence}

The pairing mechanism here depends on the stripe-like inhomogeneities,
and is expected to be stronger when the AF/stripes effects are stronger,
thus closer to the insulating side of the Mott transition regime.
Consequently one expects the pairing temperature $T_{\rm pair}$ to
decrease with the doping level $x$, as is sketched in the pairing line
in Fig.~6. 

The existence of SC requires the existence of not only pairing, but also
of phase coherence of the pairing order parameters. Under conditions
satisfied for low $x$ values, within the phase diagram of the cuprates,
pairing occurs below $T_{\rm pair}$, while SC occurs only below $T_{\rm
coh} (< T_{\rm pair})$, where phase coherence sets in. The normal-state
PG, observed in the cuprates above $T_c$ (except for high $x$ values) is
a pair-breaking gap at $T_{\rm coh} < T < T_{\rm pair}$ (see Fig.~6).
Its size and symmetry are similar to those of the SC gap, and specific
heat measurements \cite{Moca} imply that it accounts for most of the
pairing energy. The pairs in the PG state behave similarly to localized
bipolarons, and do not contribute to electrical conductivity at low 
temperatures. 

Pairing coherence requires energetic advantage of itineracy of the
pairs. Thus $T_{\rm coh}$ is expected to increase with $x$, as is
sketched in the coherence line in Fig.~6, due to moving towards the
metallic side of the Mott transition regime. Such a determination of the
pairing coherence temperature is consistent with a phenomenological
model \cite{Emery} evaluating $T_{\rm coh}$ on the basis of the phase
``stiffness''. It yields $T_{\rm coh} \propto n_s^* / m_s^*$, where
$m_s^*$ and $n_s^*$ are the effective SC pairs mass and density, in
agreement with the ``Uemura plots'' \cite{Uemura} in the PG doping
regime. 

Similarly, the existence of single-electron coherence in the normal
state, namely the existence of a Fermi liquid, depends on the advantage
of itineracy of the electronic states near $E_{_{\rm F}}$, resulting in
the increase of $T_{\rm coh}$ with $x$ (due to moving towards the
metallic side of the Mott transition regime), as is sketched in Fig.~6.
Within the non-Fermi-liquid approach used here, the stripe-like
inhomogeneities are treated adiabatically. But their dynamics becomes
faster above $T_{\rm pair}$, resulting (for $T_{\rm pair} < T < T_{\rm
coh}$) in a Fermi-liquid state where fast stripe fluctuations may still
exist. 

\begin{figure}[t] 
\begin{center}
\includegraphics[width=3.25in]{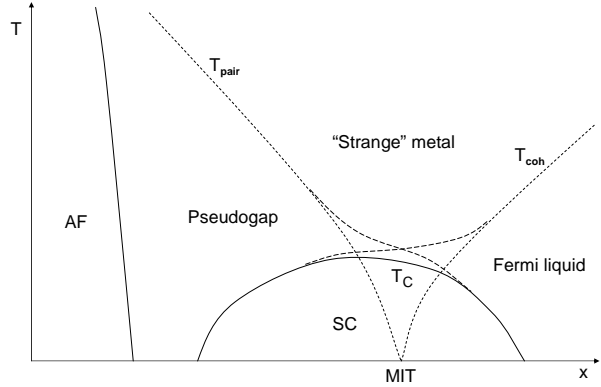}
\end{center}
\caption{A schematic phase diagram for the cuprates. The $T_c$ line is
determined by the pairing line ($T_{\rm pair}$), decreasing with $x$,
and the coherence line ($T_{\rm coh}$), increasing with $x$. Broken
lines should not be regarded as sharp lines (except when $T \to 0$), but
as crossover regimes. The MIT point is where a metal-insulator
transition occurs at $T=0$ when SC is suppressed.} 
\label{fig6}
\end{figure}

ARPES measurements in the OD regime \cite{Yusof} confirm the appearance
of coherence effects for $T_{\rm pair} < T < T_{\rm coh}$. They are
found in the nodal and antinodal peaks, and consist of the existence of
sharp peak edges and resolved bilayer-split bands. The existence of a
coherent three dimensional FS in this regime has been demonstrated
through polar angular magnetoresistance oscillations \cite{Hussey}. 

Measurements of the in-plane optical conductivity through $T_c$
\cite{marel2}, show a BCS-type behavior in the OD regime, supporting the
existence of Fermi-liquid normal state there (as shown in Fig.~6). On
the other hand, in the UD regime these measurements reveal the transfer,
below $T_c$, of spectral weight from high energies (extending over a
broad range up to at least 2~eV), to the infrared range. This behavior
has been associated with the establishment of coherence \cite{Norman}.
Within the present approach the PG state consists of localized pairs 
(and unpaired carriers) within the Hubbard gap, while the establishment
of pair coherence in the SC state requires moving states from the upper
and lower Hubbard bands into this gap, to contribute to Bloch-like
states, explaining the observed transfer of spectral weight. 

ARPES measurements on LSCO thin films \cite{Pavuna} show that $T_c$
rises under strain, while the bands close to $E_{_{\rm F}}$ become
wider. Such a change in the bands is consistent with a move towards the
metallic side of the Mott transition regime. Thus we predict that the
increase in $T_c$ under that strain is due to an increase in $T_{\rm
coh}$, and that $T_{\rm pair}$ (where the PG state sets in) may have
decreased in this case. 

As is sketched in Fig.~6, there is an increase in the values of $T_{\rm
pair}$ and $T_{\rm coh}$ in the regime where pairing and coherence
coexist, compared to their extrapolated values from the regimes where
only one of them exists. This is due to the energy gain in the SC state,
compared to both the PG and the Fermi-liquid normal states. 

If the SC state is suppressed, the occurrence of a metal-insulator
transition (MIT) is expected here at $T=0$, at the MIT point in the
phase diagram (see Fig.~6), where the metallic phase is of the
Fermi-liquid regime, and the insulating phase is of the PG regime of
localized electrons and electron pairs within the Hubbard gap. And
indeed, experiments where the SC state is suppressed by a magnetic field
\cite{Boeb}, or by doping \cite{Tallon}, show an MIT at $T \to 0$, at
$x\simeq 0.19$. This stoichiometry corresponds here to a fractional
stripon occupancy of $n^p\simeq\half$, as was determined from the TEP
results above (see Fig.~4). The existence of the MIT close to this
stoichiometry is plausible, because for higher doping levels
inter-atomic Coulomb repulsion destabilizes the dynamic charged stripes,
essential for the PG state (though the energy gain in the SC state helps
maintaining them for higher $x$). 

The MIT point in Fig.~6 is a quantum critical point (QCP), and there
have been various theoretical approaches addressing the existence of
such a QCP. Such approaches often consider different mechanisms for the
SC pairing and for the driving force of the PG state, and consequently
predict (in difference with the present approach) the existence
different regimes of different symmetries, within the SC phase
\cite{Varma}. This has not been confirmed by experiment, in agreement
with the present approach. The existence of an intrinsic nanoscale
heterogeneity of SC and PG regions \cite{Davis} will be addressed below.

The connection between SC and an MIT (where the insulating state is due
to localization) in the cuprates, as well as in other systems with a
similar phase diagram, has been pointed out by Osofsky {\it et al.}
\cite{Osofsky}. The conclusion from the present approach is that if a
pairing interaction exists, and the insulating state is characterized by
the existence of localized pairs at low temperatures, then an SC phase,
based on the same pairing interaction, ought to exist around the MIT
regime. 

\section{Pairing Gap, Excitations, and the Resonance Mode}

The QE and stripon pairing gaps $2\Delta^q$ and $2\Delta^p$ are closely
related to the order parameters $\Phi^q$ and $\Phi^p$ [see Eqs.~(15) and
(16)], and have the same symmetries. Thus $\Delta^q$ vanishes at the
nodal points and has its maximum $\Delta^q_{\rm max}$ at the antinodal
points. Since the stripons reside in about a quarter of the BZ around
$\pm{\bf k}^p$ [see Eq.~(13)], $|\Delta^p|$ does not vary much from its
mean value $\bar\Delta^p$, and it is greater than the stripon bandwidth
$\omega^p$, except for the heavily OD regime. 

Since the coupled pairing equations (17) and (18) yield in the second
order BCS-like equations for both $\Phi^q$ and $\Phi^p$, the QE and
stripon single-auxiliary-particle (Bogoliubov) energy bands in a pairing
state can be expressed as: 
\begin{eqnarray}
E^q_{\pm}({\bf k}) &=& \pm\sqrt{\bar\epsilon^q({\bf k})^2 +
\Delta^q({\bf k})^2}, \\ E^p_{\pm}({\bf k}) &=&
\pm\sqrt{\bar\epsilon^p({\bf k})^2 + \Delta^p({\bf k})^2}
\end{eqnarray}
(where $E^p_{\pm} \simeq \pm\Delta^p$ in the UD regime), and the pairing
gaps scale with $T_{\rm pair}$, approximately according to the BCS
factors, with an increase due to strong-coupling. The relevant factors
are of $s$ pairing for $\Delta^p$, and of $d$ pairing \cite{Maki} for
$\Delta^q$: 
\begin{equation}
2\bar\Delta^p \gta 3.5 k_{_{\rm B}}T_{\rm pair}, \ \ \ \ \ \ \ \
2\Delta^q_{\rm max}\gta 4.3 k_{_{\rm B}}T_{\rm pair}. 
\end{equation}

In the PG state the pairs lack phase coherence, and thus Eq.~(19) does
not yield a coherence peak in the QE gap edge. Furthermore, in this
state the low-energy svivon states are wide, due to incoherence and
scattering, and thus the gap is filled with unpaired convoluted
stripon--svivon states (see discussion below). Consequently the PG is
just a depression of width: 
\begin{equation}
2\Delta^{\rm PG}({\bf k}) = 2\Delta^q({\bf k})
\end{equation}
in the DOS, as has been observed, {\it e.g.} by tunneling
spectroscopies \cite{Renner,Kugler}. 

Pairing coherence sets in below $T_c$, and the structure of the
pair-breaking excitations is determined by the scattering between QE,
stripon, and svivon states. This scattering is strong when $E^q \simeq
E^p \pm \bar\epsilon^{\zeta}$, and particularly for svivon states close
to ${\bf k}_0$, where the $\cosh{(\xi_{{\bf k}})}$ and $\sinh{(\xi_{{\bf
k}})}$ factors, appearing in the coupling terms, are large [see Eqs.~(2)
and (3)]. When there are unpaired convoluted stripon--svivon states
within the QE gap, paired QE states are scattered to them, resulting in
the widening of the QE coherence peak [due to Eq.~(19)], at the QE gap
edge, to a hump. 

The existence of a pairing gap, and especially of the SC gap, limits the
scattering of the svivon states around ${\bf k}_0$, resulting in a
decrease in their linewidth. Let ${\bf k}_{\rm min}$ be the points of
small svivon linewidth, for which $\bar\epsilon^{\zeta}({\bf k}_{\rm
min})$ is the closest to the energy minimum $\bar\epsilon^{\zeta}({\bf
k}_0)$ (see Fig.~3). Often one has ${\bf k}_{\rm min}={\bf k}_0$, but
there are cases, like that of LSCO, where the linewidth of
$\bar\epsilon^{\zeta}$ is small not at ${\bf k}_0$, but at close points
${\bf k}_{\rm min} = {\bf k}_0 \pm {\bf q}$. The resonance mode energy
$E_{\rm res}$ is taken here as $-2 \bar\epsilon^{\zeta}({\bf k}_{\rm
min})$, accounting both for the generally observed ``commensurate mode''
at $Q = 2 {\bf k}_0$, and for cases of an ``incommensurate mode'', as
observed in LSCO at ${\bf Q} \pm 2{\bf q}$ \cite{Tran2,Wakimoto,Christ}.

The determination of ${\bf k}_{\rm min}$ is through the limitation of
the scattering of a double-svivon excitation of energy
$-2\bar\epsilon^{\zeta}({\bf k}_{\rm min})$ to a QE pair-breaking
excitation of at least the SC gap $2 |\Delta^{\rm SC}|$, by exchanging a
stripon. If $2\tilde\Delta^{\rm SC}$ (which is somewhat smaller than the
maximal SC gap $2\Delta^{\rm SC}_{\rm max}$) is the minimal energy
necessary to break a pair of QE's which are coupled to stripons around
$\pm{\bf k}^p$ through a svivon at ${\bf k}_{\rm min}$, then this
condition can be expressed as: 
\begin{equation} 
{E_{\rm res} \over 2} = |\bar\epsilon^{\zeta}({\bf k}_{\rm min})| \le
\tilde\Delta^{\rm SC}. 
\end{equation}

Consequently the svivon energies $\bar\epsilon^{\zeta}$ have a small
linewidth within the range $|\bar\epsilon^{\zeta}| \le
|\bar\epsilon^{\zeta}({\bf k}_{\rm min})|$, and for
$\bar\epsilon^{\zeta}$ within this range, the convoluted stripon--svivon
states of energies $E^p_+ \pm |\bar\epsilon^{\zeta}|$ and $E^p_- \pm
|\bar\epsilon^{\zeta}|$ form spectral peaks around
\begin{equation}
\pm E_{\rm peak}({\bf k}) = E^p_{\pm}({\bf k}\pm{\bf k}_0), 
\end{equation}
where to the ``basic'' peak width:
\begin{equation}
W_{\rm peak} = 2|\bar\epsilon^{\zeta}({\bf k}_{\rm min})| = E_{\rm res}
\end{equation}
one has to add the effects of the svivon and stripon linewidth, and of
the dispersion of $E^p_{\pm}({\bf k}\pm{\bf k}^{\prime})$ when
$\bar\epsilon^{\zeta}({\bf k}_{\rm min}) \lta \bar\epsilon^{\zeta}({\bf
k}^{\prime})\lta 0$ (see Fig.~3). The size of the SC gap is 
experimentally determined by the spacing between the closest maxima on
its two sides. Thus, in the BZ ranges around the antinodal points where
$E_{\rm peak}({\bf k})$ exists, it often determines the SC gap:
\begin{equation}
|2\Delta^{\rm SC}({\bf k})| = 2\min{[|\Delta^q({\bf k})|, E_{\rm
peak}({\bf k})]}. 
\end{equation}
[Actually, since the peak lies on the slope of the QE gap, its maximum
may be shifted to an energy slightly above $E_{\rm peak}({\bf k})$.]
$\Delta^{\rm SC}$ equals $\Delta^q$ around its zeroes at the nodal
points, which is consistent with the observation by STM \cite{Davis}
that the low energy excitations near the SC gap minimum are not affected
by heterogeneity, while the excitations near the gap edge (where it
equals $E_{\rm peak}$) are affected by it (see discussion below). Note
that there remain within the SC gap, below the peak, some unpaired
convoluted stripon--svivon states corresponding to svivon states of
large linewidth and of positive energies $\bar\epsilon^{\zeta}$ at ${\bf
k}$ points farther from ${\bf k}_0$ (see Fig.~3). 

As a result, one gets below $T_c$ a peak-dip-hump structure (on both
sides of the gap), where the peak is largely contributed by the
convoluted stripon--svivon states around $E_{\rm peak}({\bf k})$, the
dip results from the sharp descent at the upper side of this peak, and
the hump above them is of the QE gap edge and other states, widened due
to scattering to the peak states. Such a structure has been widely
observed, {\it e.g.} by tunneling measurements \cite{Renner,Kugler},
where the evolution of the gap from a depression in the PG state to a
peak-dip-hump structure in the SC state has been viewed. The appearance
of the resonance-mode energy $E_{\rm res}$ in the peak width in Eq. (25)
has been observed in tunneling measurements as the energy separation
between the SC gap edge, and the dip for different doping levels
\cite{Zasadzinski}. 

By Eqs.~(21), (22), and (26), $\Delta^{\rm PG}$ and $\Delta^{\rm SC}$
scale with $T_{\rm pair}$, and thus decrease with $x$, following the
pairing line in Fig.~6, as has been observed. Since
$-\bar\epsilon^{\zeta}({\bf k}_0)$ is zero for an AF, its value (and
thus $E_{\rm res}$) is expected to increase with $x$, distancing from an
AF state. However, by Eq.~(23) its linewidth cannot remain small if it
crosses the value of $\tilde\Delta^{\rm SC}$, which decreases with $x$.
Thus the energy $E_{\rm res}$ of a {\it sharp} resonance mode is
expected to cross over from an increase to a decrease with $x$ when it
approaches the value of $2\tilde\Delta^{\rm SC}$, as has been observed
\cite{Bourges1}. This crossover could be followed by a shift of the
resonance wave vector $2{\bf k}_{\rm min}$ from the AF wave vector ${\bf
Q}$ to incommensurate wave vectors. 

Studies of the gap structure by ARPES give information about its ${\bf
k}$ dependence, confirming the above features. In bilayer cuprates, the
QE bands are split around the antinodal points into a bonding band (BB)
and an antibonding band (AB). On the other hand, the convoluted
stripon--svivon peak given by Eqs.~(24) and (25) is {\it not} split and
extends over a range of the BZ around the antinodal points. As $x$ is
increased, $\Delta^q_{\rm max}$ is decreased, and in the OD regime it is
exceeded by $E_{\rm peak}({\bf k})$, at least in a part of the antinodal
BZ range [see Eqs.~(20) and (24)]. 

The AB lies very close to $E_{_{\rm F}}$, on the SC gap edge, and
consists of Bogoliubov quasiparticles \cite{Matsui} through the
antinodal BZ range. In the UD regime the stripon--svivon peak lies
within the QE AB gap, reflected in the observation \cite{Borisenko} of
an AB hump above the peak. On the other hand, in the OD regime the peak
lies on the QE AB, or even above it, and consequently the strong
scattering which widens the QE band edge coherence peak into a hump is
missing. Thus the reported observation is either of one AB peak
\cite{Gromko}, including the unresolved QE and stripon--svivon
contributions, or of two barely resolved peaks \cite{Feng1}, where the
QE contribution is referred to as an ``AB peak'', and the stripon--svivon
contribution, lying slightly above it, is referred to as a ``BB peak''. 

The BB, on the other hand, crosses $E_{_{\rm F}}$, and disperses up to
over $0.1\;$eV from it. Below $T_c$, when its distance from $E_{_{\rm
F}}$ is greater than that of the stripon--svivon peak, it contributes a
(QE) hump, referred to \cite{Gromko,Borisenko,Feng1} as a ``BB hump''
(though in much of this range its nature is close to that of a
normal-state band). In the range where the QE BB overlaps the
stripon--svivon peak, the fact that the electron band is formed by their
hybridized contributions results in the appearance of the antinodal kink
\cite{Gromko,Sato,Lanzara2}, due to the narrowing of the peak, as the
temperature is lowered below $T_c$. This narrowing slows down the
stripes dynamics, and widens the ``hump states'' above the peak. As has
been observed  \cite{Lanzara2}, this widening (and the resulting band
renormalization) has an isotope effect due to the significance of
lattice effects in the stripe-like inhomogeneities and in the svivon
dressing. 

The peak-dip-hump structure has been also observed in tunneling
measurements in single-layer BSCO and BSLCO \cite{Kugler}, proving that
it is not the result of just bilayer splitting. These measurements show
that in the PG state above $T_c$ in BSLCO, the stripon--svivon peak, and
the QE hump are merged into one hump. ARPES results in the PG state of
BSLCO \cite{Janowitz} show an apparent ``bilayer splitting'' which may
indicate that even though the QE and the stripon--svivon contributions
to this hump appear merged in tunneling results, they are separated from
each other in different ${\bf k}$ points. 

In the heavily OD regime, where $\Delta^{\rm SC}$ and $E_{\rm res}$
become smaller than $\omega^p$, the dispersion of $E_{\rm peak}({\bf k})$
becomes wider than $\Delta^{\rm SC}$ and $W_{\rm peak}$ [see Eqs.~(20),
(24) and (25)]. This results in the smearing of the peak in
spectroscopies where it is integrated over the BZ, though it may still
be detected in different ${\bf k}$ points [where there remains some 
smearing, as was mentioned below Eq.~(25)]. The apparent disappearance,
in the heavily OD regime, of the peak (of width $E_{\rm res}$) observed
in optical measurements \cite{Timusk} below $T_c$ may be the result of
such smearing (see Ref.~\cite{Cuk}). Concerning the question
\cite{Timusk} whether the resonance mode is significant for high-$T_c$
SC, the approach presented here considers svivons in the vicinity of
${\bf k}_0$ to be significant for the SC pairing, whether they
contribute to the narrow resonance mode peak, or to higher energy
excitations \cite{Tran3,Hayden}. 

\section{Heterogeneity and Pairs Density in the SC Phase}

The narrow stripon band splits in the SC state, through
the Bogoliubov transformation, into the $E^p_{-}({\bf k})$ and
$E^p_{+}({\bf k})$ bands, given in Eq.~(20). The states in these bands
are created, respectively, by $p^{\dagger}_{-}({\bf k})$ and
$p^{\dagger}_{+}({\bf k})$, which are expressed in terms of creation and
annihilation operators of stripons of the two pairing subsets [see
Eq.~(16)] through equations of the form: 
\begin{eqnarray}
p^{\dagg}_{-}({\bf k}) &=& u_{\bf k} p^{\dagg}_{\bigtriangleup}({\bf k})
+ v_{\bf k} p^{\dagger}_{\bigtriangledown}(-{\bf k}), \nonumber \\
p^{\dagger}_{+}({\bf k}) &=& -v_{\bf k}
p^{\dagger}_{\bigtriangleup}({\bf k}) + u_{\bf k}
p^{\dagg}_{\bigtriangledown}(-{\bf k}), 
\end{eqnarray} 
where $|u_{\bf k}|^2 + |v_{\bf k}|^2 = 1$. 

If all the stripons were paired, then at low temperatures, where the
$E^p_{-}$ band is completely full, and the $E^p_{+}$ band empty, then
the fractional stripon occupancy $n^p$ should have been equal to
$\langle |u_{\bf k}|^2 \rangle$. However this cannot be fulfilled in the
UD regime, where $\bar\Delta^p$, is considerably greater than the
stripon bandwidth $\omega^p$, and $E^p_{\pm} \simeq \pm\Delta^p$,
resulting in $\langle |u_{\bf k}|^2 \rangle \simeq \half$, while $n^p >
\half$ (see Fig.~4). 

Consequently, in the UD regime, the PG state should consist of {\it
both} paired and unpaired stripons (as was discussed above), and the SC
phase should be {\it intrinsically} heterogenous with nanoscale SC
regions, where, locally, $n^p \simeq \half$, and PG regions where,
locally $n^p > \half$, such that the correct average $n^p$ for that
stoichiometry is obtained. The size of the regions in this nanostructure
should be as small as permitted by the coherence length, and it was
indeed observed in nanoscale tunneling measurements in the SC phase
\cite{Davis}. There are, however, physical properties which are
determined through a larger scale averaging over these regions. This
nanostructure would be naturally pinned to defects, and could become
dynamic in very ``clean'' crystals. 

For $x\simeq 0.19$, one has $n^p \simeq \half$, and an SC phase could
exist without such a nanostructure. Furthermore, in the OD regime
$\bar\Delta^p$ becomes comparable, and even smaller than $\omega^p$, and
the condition $\langle |u_{\bf k}|^2 \rangle = n^p$ could be satisfied
with all the stripons being paired. Thus the SC phase could exist in the
OD regime (especially for $x \gta 0.19$) without the above nanoscale
heterogeneity, as has been observed \cite{Davis}. 

The Uemura plots \cite{Uemura} give information about the effective
density of SC pairs $n_s^*$. Within the present approach, the pair
states are fluctuating between QE and stripon pair states, and thus
$n_s^*$ is determined by the smaller one, {\it i.e.}, the density of
stripon pairs. As was discussed above, the stripon band is half full for
$x\simeq 0.19$, and consequently $n_s^*$ should be maximal around this
stoichiometry, being determined (for p-type cuprates) by the density of
hole-like stripon pairs for $x\lta 0.19$, and of particle-like stripon
pairs for $x\gta 0.19$. This result is not changed by the intrinsic
heterogeneity for $x\lta 0.19$, since even though the stripon band
remains approximately half full within the SC regions, the fraction of
space covered by these regions, and thus $n_s^*$, is {\it increasing}
with $x$ in this regime, while in the $x\gta 0.19$ regime, $n_s^*$ is
{\it decreasing} with $x$ because the occupation of stripon band is
decreasing below half filling. This result is consistent with the
``boomerang-type'' behavior \cite{Niedermayer} of the Uemura plots
around $x\simeq 0.19$. 

Low temperature ARPES results for the spectral weight within the SC peak
(omitting the background including the hump), integrated over the
antinodal BZ area \cite{Feng2}, reveal a maximum for $x\simeq 0.19$
(similarly to $n_s^*$). This is expected here, assuming that the
integrated spectral weight counted is dominantly within the
stripon--svivon peak (discussed above), and that this peak counts the
major part of SC hole-like pair-breaking excitations of stripons within
the $E^p_{-}$ band, as is expected. For the intrinsically heterogenous
$x\lta 0.19$ regime, the integrated weight within the peak is expected
to increase with $x$ because of the increase in the fraction of space
covered by the SC regions (keeping the stripon band approximately half
full within these region). For the $x\gta 0.19$ regime the integrated
weight, measured by ARPES, scales with $\langle |u_{\bf k}|^2 \rangle$,
and is thus decreasing with $x$ below half filling of the stripon band. 
Note that the contribution of the QE AB to the ARPES peak had to be
omitted \cite{Feng2} in order to get the decrease of the peak weight for
$x\gta 0.19$, confirming that this behavior is due to the
stripon--svivon peak, as is suggested here. 

\section{Conclusions}

The anomalous properties of the cuprates, including the occurrence of
high-$T_c$ superconductivity, are found to be a result typical of their
electronic and lattice structure, within the regime of a Mott
transition. On one hand, hopping-induced pairing, which depends on
dynamical stripe-like inhomogeneities, is stronger for low doping
levels, closer to the insulating side of the Mott transition regime. On
the other hand, phase coherence, which is necessary for
superconductivity to occur, is stronger for high doping levels, closer
to the metallic side of the Mott transition regime. Pairing without
coherence results in the pseudogap state of localized electrons and
electron pairs, and coherence without pairing results in a Fermi-liquid
normal state. Suppression of superconductivity results in a quantum
critical point of a metal-insulator transition between these two states
at $T=0$. An intrinsic heterogeneity exists of nanoscale superconducting
and pseudogap regions. 
 

\end{document}